\documentclass{elsart}
\usepackage{natbib}
\usepackage{graphicx}
\usepackage{amssymb}
\begin{document}
\begin{frontmatter}
\title{Pulsar Magnetic Field Oscillation Model and Verification Methods}
\author{Zhu-Xing Liang}
\ead{zx.liang55@hotmail.com}
\author{Yi Liang}
\ead{jluliangyi@gmail.com}
\address{18-4-102 Shuixiehuadu, Zhufengdajie, Shijiazhuang, Hebei 050035,China}

\begin{abstract}
We constructed the magnetic field oscillation model (hereafter the MO model) by analogizing the periodically reversing phenomenon of the solar magnetic field to pulsars. Almost all kinds of pulsar radiation phenomena are best explained using the MO model, especially polarization characteristics, glitch, generation rate, the geodetic precession of pulsars and the configuration of pulsar-wind nebula of the Crab. The MO model also provides satisfactory explanation for other characteristics of pulsars, e.g., interpulse, spin-down, pulse nulling, beat and pulse drift, the loss rate of the rotating energy, and the accuracy of frequency. We present eight verification methods for the MO model. In addition to pulsars, our MO model can also be used to explain the pulse emission from non-compact stars such as the ultracool dwarf TVLM 513-46546 and the magnetic chemically peculiar star CU Virginis.

\end{abstract}
\begin{keyword}
Pulsars \sep Oscillations \sep  Radiation mechanisms \sep
Polarization

\PACS  97.60.Gb \sep 97.10.Sj \sep 95.30.Gv
\end{keyword}
\end{frontmatter}

\newpage

\tableofcontents

\section{Introduction}
\label{} In studies of pulsars, three models have been discussed in
detail: the oscillation, the orbiting binary system, and the
lighthouse model. Once the oscillation and the orbiting binary
system were ruled out on the basis of calculation results, the
lighthouse model came to be widely accepted. The idea of analogizing
the reversing phenomenon of the solar magnetic field to magnetic
variables \citep{Kienle50, Babcock58} has been proposed previously;
however, it hasn't been taken seriously in the study of pulsars. Our
study demonstrates that the magnetic field oscillation model
(hereafter MO model) can be constructed by analogizing the reversing
phenomenon of solar magnetic field to pulsars, and that the MO model
provides a satisfactory explanation of the radiation characteristics
of pulsars. This paper comprises our research results together with
six methods for testing the validity of the MO model.

This paper explains the various observed characteristics of pulsars
using the MO model. The reasons why the magnetic fields of pulsars
reverse rapidly are beyond the scope of this paper, and will be
considered in a separate paper.
\section{MO model}
We know that the solar general magnetic field reverses every 11
years, representing an oscillating period of 22 years. The MO model
considers that the pulse radiation of pulsars originates from
oscillations of the magnetic field, just similar to that of the Sun.

The MO model believes that the magnetic field of pulsars oscillates
in a similar way to that of the sun, although with a much higher
frequency. The rate of the magnetic flux change is very high during
reversals of the magnetic field. Therefore, a very high ring-shaped
induction voltage is generated during this period. If this induction
voltage is calculated using the sine law, the peak value of the
induction $E_{max} =  \pi Bd / 2P$. Supposing that the diameter of
the pulsar is $d = 10^{4}$ m, the oscillation period is $P=1$ s, the
peak value of the magnetic field within the pulsar is $B = 10^{8}$
T, the induction electrofield intensity on the surface of the pulsar
is $E_{max}= 1.56 \times 10^{12}$ V/m. If the change of the magnetic
field is similar to a square wave, the electrofield intensity will
be much higher.

\begin{figure}
  \includegraphics[width=12cm]{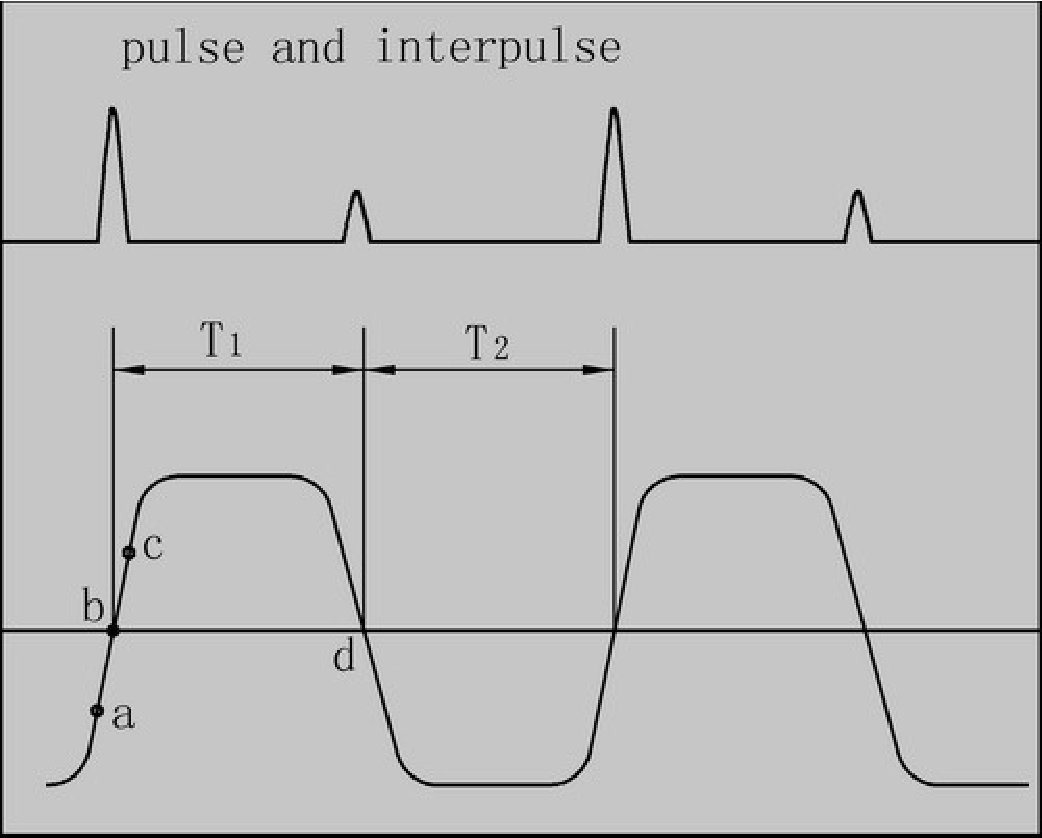}\\
  \caption{Magnetic field oscillation and pulse radiation}\label{f1}
\end{figure}

The effect of induction voltage means that the charged particles
around pulsars are accelerated along the circle until approaching
the speed of light, at which point sync radiation is generated.
Because the radiation only occurs in some moments (for example, near
points b and d in Figure \ref{f1} when the magnetic field changes
faster), the observed radiation is shown in the form of pulse. In a
period, only one or two of the strongest pulses can usually be
received. However, it cannot be eliminated that more weak components
appear, such as the Crab pulsar on which there are five peaks to be
observed \citep{MH96}.

The effect of induction voltage means that the pulse radiation is
also potentially generated on the surface of pulsars, similar to the
process of the generation of lightning.

The sun is a common main-sequence star. As the solar magnetic field
oscillates, other main-sequence stars should also oscillate. In
addition, the earth, although markedly different from the sun, also
shows periodic reversals in its magnetic field. These observations
lead us to believe that oscillations in the reversals of magnetic
fields are a shared characteristic of all celestial bodies,
including planets, main-sequence stars, white dwarfs and pulsars.
The characteristics of the magnetic field oscillation of pulsars is
nothing more than an extremely short time-scale of oscillation
resulting from extreme physical conditions.

The discovery of the ultracool dwarf radiation \citep
{Hallinan07} also indicates that the pulse radiation is not peculiar
to neutron stars.
A fundamental assertion of the MO model is that the radiation pulses (at least narrow pulses)
always appear close to the zero point of magnetic field intensity. It is noticeable that the observation of
\citet{Trigilio00} indicates that this is just the case for CU Virginis. We think that the result is not pure coincidence.
For these reasons, we guess that the radiation and
the photometric change of all kind of celestial bodies are driven by
the reversal oscillation of their magnetic fields.

After analyzing,
we've found there is no strong evidence to negate our guess, but
there are many evidences to support us.

\begin{table}[htbp]
  \caption{The oscillation time-scales of different Celestial body}
  \label{tab:1}
  \begin{tabular*}{\hsize}{ll}
\hline
Celestial body & Oscillation time-scale\\
\hline
earth     & about 500,000 years\\
sun    & about 22 years\\
non-compact star   & several hours to several decades\\
white dwarf  & several seconds to several minutes\\
pulsar  & sub-millisecond to several seconds\\
\hline
\end{tabular*}
\end{table}

Table \ref{tab:1} shows the oscillation time-scales of different
celestial bodies.

We don't yet have a satisfactory understanding of why the magnetic
fields of stars are able to oscillate. Nevertheless, the following
discussion shows that the observed characteristics of pulsars are
able to sufficiently explain their oscillatory nature.
\section{Discussion}

\subsection{Polarization phenomenon }
\begin{figure}
  \includegraphics[width=12cm]{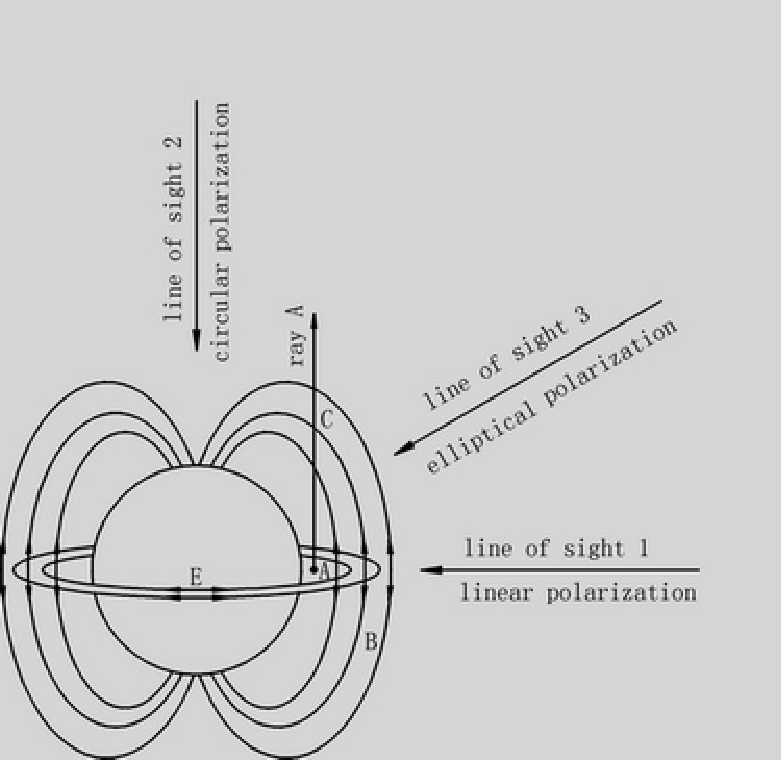}\\
  \caption{Relationship between the spin axis, line of sight
  and polarization characteristics}\label{polarization}
\end{figure}

It is very difficult to use the lighthouse model to explain the
reversion of rotation direction in some circular polarizations. In
contrast, the MO model provides a sound explanation of such
phenomena, as shown in Figure \ref{polarization}. According to the
MO model, the accelerated particles revolve around pulsars along the
latitude. As we know, these particles are able to radiate circularly
polarized wave along the axial direction and as a linearly polarized
wave along the plane of the equator. If our line of sight is
superposed with the axial line, the circular polarization can be
seen, whereas if our line of sight is perpendicular to the axis, the
linear polarization can be seen. If the line of sight is neither
perpendicular nor parallel to the axis, the elliptical polarization
can be seen. Therefore, the polarization characteristic depends on
the angle between the spin axis and the line of sight. Because the
rays must pass through the pulsar's external magnetosphere when
traveling away from the pulsar, the light wave generates Faraday
rotation, leading to a change in the polarization parameter. The
time at which the rays pass through the magnetosphere is the period
when the magnetic field reverses. At this time, the intensity of the
magnetic field is alternately weakening and strengthening.
Therefore, Faraday rotation changes the position angle of the linear
polarization to an S form. The rotation direction can also reverse
in the circular polarization. Of course, the magnetic field does not
vary exactly according to sine, and it is not certain that the
radiation appears exactly when the magnetic field reverses through
the zero point. Accordingly, the actual change in the polarization
is much more complicated.

As shown in Figure \ref{polarization}, the radiation of the linear
polarization can retain its high linear polarization degree when the
ray passes through the magnetosphere transversally, though its
position angle changes in the S form. However, when ray A in region
C passes through the magnetosphere transversally, the anisotropic
properties can change the circular polarization into elliptical
polarization. Therefore, the observed circular polarization degree
is always lower.

\subsection{Configuration of pulsar-wind nebula }
\begin{figure}
  \includegraphics[width=9cm]{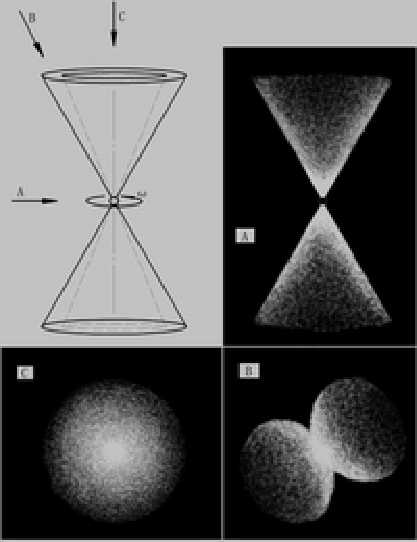}\\
  \caption{Configuration of the PWN}\label{4 panel}
\end{figure}

If there is a pulsar within a supernova remnant, it will be possible
to view a pulsar-wind nebula(PWN) using the Chandra telescope. The
lighthouse model states that the radiation particles of pulsars are
distributed in a configuration that resembles two cones placed tip
to tip, as shown in the upper-left panel of Figure \ref{4 panel}.
When this structure is observed from direction B, not only the
hyperbola bright faculae shown in panel B can be seen, but the pulse
signal from pulsars can  also be received. If observed from
directions A or C, only the PWNs can be seen, as in panels A and C
in Figure \ref{4 panel}; no pulse signal is received because the
beams of light do not sweep over the earth in these two cases.

The picture of the supernova remnant taken by the Chandra telescope
does not support the conjecture of the lighthouse model.

\textbf{3.2.1} In all of the remnants within which pulsars have been
found, such as Crab and Vale, the bright hyperbola faculae shown in
panel B of Figure \ref{4 panel} has not yet to be found.

\textbf{3.2.2} In all of remnants within which no pulsars have been
found, the faculae shown in panels A and C of Figure \ref{4 panel}
have not yet to be found either.

\begin{figure}
  \includegraphics[width=9cm]{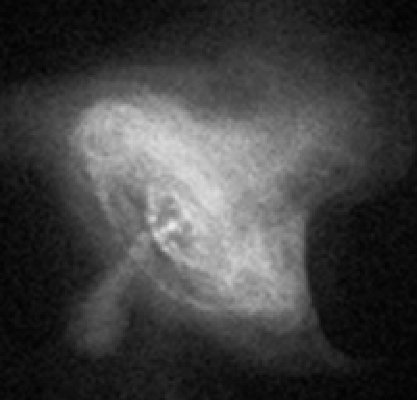}\\
  \caption{PWN of Crab \citep{Weisskopf00}} \label{Crab}
\end{figure}

According to the MO model, during reversal of the magnetic field the
speed of the particles is extremely high. When the particles are
accelerated circularly within the equatorial plane, they move
outward under the influence of inertial force, while the magnetic
field goes through zero point with a very weak intensity.
Accordingly, a rotiform particle distribution forms around pulsars.
In addition, the MO model predicts that the magnetic axis of pulsars
always basically aligns with its spin axis. Therefore, trumpetshaped
magnetic lines should occur near the poles, and some particles
should flow out from the polar region along the magnetic lines to
form an axial particle distribution. Figure \ref{Crab} shows a
distinct rotiform and axial configuration, and the pulsar is located
in the center of the nebula disk rather than at the apex of two
coins. This image strongly supports the validity of the MO model.

The MO model can well explain the filaments and the annular
structures in PWN. The particles around a pulsar move outward by the
action of an annular induced electric field. Where the particles are
dense, each particle obtains less energy and the moving speed is
lower, whereas, where the particles are thin, it can get more energy
and the moving speed is higher. That is, the denser the particles,
the lower the speed; the thinner the particles, the higher the
speed. In this way, the thin particles chase the dense particles and
accumulate to form a lot of filaments. Then, the filaments
accumulate to form denser annular structure. The filaments and the
annular structures are very clear in Crab and Vela nebulas.

\subsection{Frequency and stability of the frequency}
At first glance, it is almost unimaginable that the reversal
frequency of the pulsar magnetic field is so high; however, as long
as we consider that the reversal period shows a positive correlative
with the stellar capacitance and induction capacity, and a negative
correlative with the rotating speed and temperature, then a
millisecond-level oscillating period is plausible for pulsars.

Our proposal is that the factors that control reversals of the
stellar magnetic field are the stellar components, geometric
dimension, rotating frequency, temperature, and charge radiant rate,
rather than the very random behavior of eddies. Because these
factors have very high long-term stability, the reversal frequency
of the stellar magnetic field must have a very high long-term
stability. If one insists on explaining the origin and reversal of
the magnetic field using eddy-current generator theory, it would be
problematic to explain both the frequency stability of the pulsars
and the frequency stability of the solar cycle.

On earth, the fundamental reason for the lower frequency accuracy of
all kinds of electromagnetic oscillations is that there is no way to
completely eliminate environmental interference. On pulsars, there
is no such environmental interference at all: they are all isolated
systems within empty space. Therefore, it is normal for pulsars to
have a very high accuracy in frequency. The accuracy of the sun's
oscillation is possibly degraded by nuclear reactions that take
place within the sun. In the absence of nuclear reactions within
pulsars, the oscillation of magnetic field to be powered by only
rotating energy means that a very high accuracy is easily
maintained.

It is important to ask why the accuracy of the solar magnetic field
period is lower than that of pulsars. One of the reasons is that
there are just 11 periods (22 half- periods) in which to record
oscillations in the solar magnetic field. If only 11 signal periods
of a pulsar are considered, the resulting period accuracy is also
very low, probably approximately the same as that of the solar
period. This indicates that there is no clear difference between the
short-term stability of the sun and that of pulsars. There is
presently no way of knowing the long-term stability of the sun. Once
we are able to make billions of records of solar periods, we will be
able to calculate the long-term stability of the sun and the solar
periods to a very high degree of accuracy.

There is no comparability between the short-term stability of the
solar period and the long-term stability of a pulsar period.
Therefore, the validity of the MO model can be assessed by comparing
these two features.

\subsection{Glitch }
The lighthouse model considers that the glitch is the result of a
sudden change in the period of rotation. The model provides three
explanations of why the period of rotation changes, but problems
remain with these explanations.

\textbf{Impact of extraneous objects.}  If the change in the period
of rotation is considered to result from the impact of extraneous
objects, there is no way of explaining the recovery of the rotating
velocity after it has sped up.

\textbf{Change in the rotating inertia.} If it is considered that
the change in the period of rotation results from the change in
rotating inertia, we must consider (1) why the rotating velocity
always increases first and then decreases, (2) why it always
increases rapidly but decreases slowly, and (3) why no adverse
phenomenon takes place. It is very difficult to explain these
questions. In addition, before and after the glitch, the moment of
momentum is conservative, yet the rotating energy is not
conservative. It is necessary to transform other energy into the
rotating energy. During the slow recovery process, the increased
part of the rotating energy has to be completely transformed into
other forms of energy. It is very difficult to associate this
reversible energy transformation with a catastrophic event such as
an impact or starquake.

\textbf{Exchange of momentum between the inner matter and the outer
shell.} If the exchange of momentum between the inner matter of
pulsars and the outer shell cause periodic glitch, there must be a
differential rotation between the inner and outer matter. However,
differential rotation contradicts the fixed inclination, unless the
magnetic field is only installed on the thin shell and has nothing
to do with the inner matter, the lighthouse model cannot justify
itself.

In contrast to the lighthouse model, the MO model provides a much
more simple explanation of the glitch. The MO model considers that
the frequency of pulsar oscillation is most closely related to
atmospheric temperature. When the temperature rises, the frequency
also increases. The main cause of the period glitch is the sudden
change in the magnetospheric temperature of pulsars, mainly caused
by the impact of extraneous objects. These catastrophic events
result in a rapid increase in temperature followed by a slow
decline. Accordingly, the oscillation frequency always increases
rapidly and then decreases slowly. The recovery speed of the
frequency depends on the recovery speed of the temperature. If more
heat is produced in the atmosphere and the heat diffuses into the
inner part of the pulsar, a clear change occurs in the temperature
of the celestial body and the frequency will not recover completely.

The collision of extraneous objects can lead to a change in the
rotating period with increasing temperature. This change generally
features a sudden increase rather than a slowing down. Because most
extraneous objects are the outer substances of pre-stars, as with
the planets of the solar system, their rotating direction is
identical to the direction of pulsar rotation. Therefore, collision
will inevitably accelerate the rotation rate of a pulsar. It is
hypothesized that the oscillation frequency shows a positive
correlation with the rotating frequency. In this case, the collision
will first cause the rotation velocity to suddenly increase, and
then the sudden increase of the rotation velocity results in a
sudden increase of the oscillating frequency. This sudden increase
in oscillating frequency cannot recover. That is, the glitch caused
by the collision can be divided into two parts: a recoverable part
and an irrecoverable one.

Starquakes can also cause glitch; however, in this case it first
leads to a direct and immediate change in both the rotating period
and temperature, followed by an indirect change in the oscillating
frequency via the changes in rotation and temperature.

\subsection{Birth-rate of pulsars }
Observations of the Galaxy indicate that a supernova outburst is
unable to create a large number of pulsars. The difference between
estimates of the lighthouse model and observations is one digit. If
the MO model is accepted, there is no such difference because the MO
model does not recognize those pulsars that are concealed because
their beams of light do not sweep over the earth. Therefore, the
number of pulsars estimated by the MO model is far smaller than that
estimated by the lighthouse model. The number of pulsars estimated
by the MO model is more approximate to the number actually observed.

\subsection{Spin down and age }
The fact that the pulsar period increases over time and that
calculated ages are consistent with observed ages is considered to
represent solid evidence of the validity of the lighthouse model;
however, as long as we consider that the oscillating frequency of
the magnetic field shows a positive correlation with its rotating
velocity, these observations are also entirely consistent with the
MO model.

\subsection{Pulse nulling, beat and pulse drift}
Some pulsars have pulse nulling, beat, or pulse drift. The MO model
provides two explanations for the occurrences of these phenomena.

The first reason is the change of the magnetosphere environment. The
condition of the magnetosphere is not uniform, with structures
similar to sunspots drifting within the magnetosphere. This
heterogeneity can lead to continuous changes in the electromagnetic
environment of the sparking region. At times, when it is easy to
spark, the pulse will appear early; however, at times when it is not
easy to spark the pulse will appear a little late. The continuous
change in the electromagnetic environment can lead to a change in
the strength of the single pulse. The pulse fluctuation is
occasionally strong enough to have nulling. The resulting pulse
fluctuation is largely random. Nevertheless, if the period of the
change in electromagnetic environment is close to an integral
multiple of the pulse period, phenomena termed beat or pulse drift
will appear. The frequency of the pulse will directly reflect the
oscillating frequency of the magnetic field, and the periods of the
beat or the pulse drift will provide indirect information on changes
in the period of the electromagnetic environment.

The second reason is the defilade function of pulsars. If the pulsar
has a hard shell and anomalous structures such as hills on its
surface, radiation will occur at places where it is easy to spark.
If the sparking spots are fixed to smaller regions, they will be
obscured from sight when the pulsar rotates and the regions are
located on the back of the celestial body; pulse nulling will then
appear. If the sparking region is larger than 180$^{\circ}$ in
longitude, only the fluctuation in the pulse strength occurs, with
no defilade phenomenon. The pulse frequency of pulsars shows a
positive correlation with spin frequency, although the two are
generally not equal and sometimes they can be significantly
different. If the pulse frequency happens to be close to an integral
multiple of the pulsar rotating frequency, beat phenomenon will also
appear.

It is noteworthy that beat or pulses drift is a typical modulation
phenomenon between waves. When two waves with different frequencies
modulate, if one period is close to another, the drift will appear;
if the difference of two periods is an integral multiple, beat will
appear. However, only a few pulsars show the clear modulation
feature and the spin modulating effect might be very slight for the
most of pulsars. A significant example for slight modulation is the
1/60 Hz signal in Crab \citep{Cadez01}. We insist that the 1/60 Hz
signal is the spin signal, though it has been explained as free
precession effect.

Beat and pulse drift demonstrate that there are at least two
periodic phenomena with different frequencies on a pulsar; however,
a basic hypothesis of the lighthouse model is that all of the
substances and magnetic field on pulsars should co-rotate with the
stellar body. According to this hypothesis, there is only one
frequency: spin frequency. Nevertheless, a second hypothesis that is
invoked when the lighthouse model explains drift is that particles
in the magnetosphere revolve relative to the magnetic axis. These
two hypotheses are clearly contradictory. As the MO model does not
require a hypothesis of magnetosphere co-rotation, no
self-contradiction occurs.

\subsection{Rate of energy loss in rotating }
According to the MO model, the factor that controls particle
acceleration is the rate of change of the magnetic flux, rather than
the strength of magnetic induction. When the magnetic field density
changes through the zero point, the majority of the magnetic energy
changes into electric field energy to accelerate the particles; the
rate of energy utilization is much greater than that described in
the lighthouse model. It is therefore possible that the intensity of
the pulsar magnetic field is much weaker than that predicted by the
lighthouse model. During reversal of the magnetic pole, the magnetic
field energy first changes into electric field energy, with part of
the electric field energy being radiated out or taken away by
particles before most of the electric field energy changes back into
magnetic field energy. Therefore, only a small part of the energy
radiated out or taken away is needed to be supplemented with the
rotating energy.

In addition, the MO model doesn't recognize dipole radiation, and
the previous method used to estimate the magnetic field intensity no
longer has any effect. After taking these changes into account, it
is necessary to use a new method to estimate the rate of the loss of
rotating energy. During the time that it takes a new estimation
method to be formulated, the question of the rate of the loss of
rotating energy is not a sound or sufficient reason to reject the MO
model.

\subsection{Interpulse }
It is difficult to explain the interpulse using the lighthouse
model, especially the interpulse of the Crab pulsar. From Figure
\ref{Crab}, the angle between the spin axis and our line of sight is
estimated to be about 60$^{\circ}$. According to the lighthouse
model, it should have no interpulses at this angle. Howevere, the
Crab pulsar does have an interpulse. Therefore, the image of the
Crab pulsar does not support the validity of the lighthouse model.

According to the MO model, the Lorentz force generated by pulsar
rotation can force the electrons to reciprocate along with radius
direction. This reciprocating motion can influence its
electromagnetic environment anywhere within or outside of a pulsar.
Periodic variations in sunspots reflect periodic changes in the
electromagnetic environment. If similar variations in
electromagnetic environment occur on pulsars, the pulse radiation
parameter would be affected. Therefore, the parameter at point b in
Figure \ref{f1} would differ from that at point d. In other words,
the parameter of the odd pulse differs from that of the even pulse;
this is the odd-even difference. In addition, if T$_{1}$ is not
strictly equal to T$_{2}$, or if the slope of the magnetic flux
variation at point b is not equal to the slope at point d, the
odd-even difference is also generated. The odd-even difference has
four kinds of representations: peak value difference, phase
difference, shape difference, and polarization difference.

As we known, on average, sunspot numbers in odd periods are larger
than those in even periods. If the record curve of sunspots is
folded taking a period of 22 years, an interpulse appears that is
similar to those of pulsars. This indicates that when the magnetic
fields of stars oscillate and reverse, there will be odd-even
differences in the observed data. If the radiation of pulsars
originates from oscillations in the magnetic field, we would then
have a satisfactory explanation for the interpulse.

The MO model predicts that the odd-even difference exists in any
pulsar radiation. Nevertheless, it is true for most pulsars that the
odd-even difference is so large that only the strongest interpulse
is able to be observed. The MO model predicts that if the
signal-to-noise ratio can be greatly increased, many new examples of
interpulses are sure to be found.

Owing to interpulses mostly appearing on the young pulsars, we guess
that the negative charges can be accelerated much more easily than
the positive ones escaping from celestial body, it can result in the
superabundance of positive charges and a departure from electrical
neutrality on the old pulsars, consequently, so great odd-even
difference would occur that the interpulse would be too small to be
seen. In other words, it is the excursion degree of electrical
neutrality that controls that the odd-even difference is great or
small and whether the interpulse can be detected .

\subsection{Micropulse }
It is difficult to explain the fine structure of a pulse with
reference to the lighthouse model. Assuming that the fine structure
of each pulse corresponds to the fine structure of the radiation
region, our calculations show that even if completely coherent
radiation is generated in all of the fine regions, just as would
occur if the array was composed of many lasers, it is impossible to
generate such a fine micro-pulse.

The scattering law of laser beams is
\begin{equation}
\tan\beta=\frac{2\lambda}{\pi d} , \label{e1}
\end{equation}
where $\beta$ is the divergence angle, $\lambda$ is the wavelength,
and $d$ is the diameter of the emission region.

According to equation (\ref{e1}), the smaller the beam diameter of
the emission region, the larger the divergence angle. When the beam
divergence angle $\beta$ and its wavelength $\lambda$ are known, the
lower limit of the geometry dimension of the emission source can be
calculated.

Taking PSR B1133+16 as an example, its period is $1.188$ s. At the
1.65 GHz waveband, the narrowest micro-pulse width is 2 $\mu $s
\citep{Popov02}, and the ratio of the micropulse width to the pulse
period is $1.68 \times 10^{-6}$; therefore, the beam divergence
angle is limited to $\beta \leq 6.06 \times 10^{-4}$ degree. If the
micropulses correspond to microstructures of the radiation source,
the diameter of the microstructures is $d \geq 10.8 $ km, which is
impossible.

Special attention is drawn to the fact that the calculation above is
based on completely coherent radiations. In fact, it is impossible
to have completely coherent radiations in the emission region
because only homogeneous light can be coherent, whereas pulsars
radiate according to the power-law spectrum. For incoherent
radiation, the calculation should be performed according to the area
light source. Therefore, the beam divergence angle is much larger
than that given above, which is not at all likely to have a
micropulse with a very short time-scale.

According to the MO model, the signal of pulsars is a real
time-domain signal, as with lightning radiation on earth. The
characteristic of this kind of signal is that the greater the
propagation distance, the weaker the signal, but the fine structure
of a pulse is retained in its entirety.

\subsection{Acceleration of particles }
According to the lighthouse model, the charged particles are
accelerated in the inner gap or the outer gap; however, there is
considerable doubt concerning the presence of the so-called inner
gap or outer gap. It is known that particle movement in the
accelerating region neutralizes the original accelerating field. To
maintain the strength of the electro field, it is necessary to
supplement it with energy. The supplementary energy must be equal to
that taken away by the particles. The process of the energy
transformation is as follows: stellar kinetic energy $\Rightarrow$
magnetic energy $\Rightarrow$ potential energy of the accelerating
electro field $\Rightarrow$ particle kinetic energy. In these three
steps, it is considerably doubtful as to whether the second step of
the energy transformation is possible, as it is very difficult to
explain how the energy flows continuously into and out of the small
space near the magnetic pole.

According to the MO model, during reversal of the magnetic field the
induced electromotive force is generated around pulsars. This
induced electromotive force is the power required to accelerate the
particles. The process of the energy transformation is as follows:
stellar kinetic energy $\Rightarrow$ magnetic energy $\Rightarrow$
circle-induced electro field potential energy $\Rightarrow$ particle
kinetic energy. Every step in this process is certainly possible;
oscillation of the solar magnetic field is completely consistent
with this process although its frequency is very low. In terms of
explaining particle acceleration and the energy supplement, the MO
model is more credible than the lighthouse model.

\subsection{Magnetic inclination }
A basic hypothesis of the lighthouse model is that the magnetic
inclination is non-zero; however, a simple analysis demonstrates
that the magnetic inclination of a neutron star generated from a
supernova explosion is equal to zero.

Before a supernova explosion, the density of inner matter within the
pre-star is higher than that of the outer matter. During the
explosion process, the matter of the star collapses inwards the
center. As the outer matter is less dense, the contraction rate in
the radius direction is higher. According to the conservation of
angular momentum, after collapsing the outer angular velocity is
higher than the inner one. This relative rotation between different
layers means that the magnetic inclination eventually reaches zero.

The hypothesis that the inclination is non-zero requires that the
magnetic field is inserted onto the solid, and that the attached
body of the magnetic field does not have differential rotation;
otherwise, the magnetic inclination would disappear.

The fact that the magnetic inclination is incompatible with the
differential rotation is a problem for the lighthouse model because
pulsars are commonly considered to have a thin shell that contains
liquid, and it is therefore difficult to avoid differential
rotation. The MO model is not affected by this problem, as it
considers that the magnetic inclination is equal to zero.

\subsection{Polar motion of pulsars }
\citet{Monaghan68} and several other researchers studying magnetic
stars once stated that magnetic dipole radiation can cause the
magnetic poles to move.In fact, this is polar motion and it is
suitable for pulsars.

For pulsars, the change law of the magnetic inclination is
\begin{equation}
\dot{\theta}=\frac{\dot{P}}{P\tan\theta}, \label{e2}
\end{equation}
or
\begin{equation}
\dot{\theta}=\frac{1}{2T\tan\theta}, \label{e3}
\end{equation}
where $\theta$  is the magnetic inclination, $ \dot{\theta}$ is the
derivative of $\theta$,  $\dot{P}$ is the derivative of period and
$T$ is the characteristic age.

The equations (\ref{e2}) and (\ref{e3}) show that the magnetic
inclination decreases quite rapidly when a pulsar has both a small
characteristic age and a small magnetic inclination.

For example, Crab pulsar's characteristic age is about 1,240 years,
and the angle between the spin axis and our line of sight is
$60^{\circ}$ \citep{Weisskopf00}. We are sure that the magnetic
inclination is also $60^{\circ}$, otherwise, we can not see its
light beam. Substituting these data into equation (\ref{e3}), we
find that the magnetic inclination should decrease by about
$0.013^{\circ}$ per year. If this result is authentic, the life of
Crab pulsar must be less than several thousand years. If the result
were not authentic, the lighthouse model which is the calculating
base might be also not authentic.

The MO model considers that the moment of momentum of pulsars can be
transferred to the outer mass by electromagnetic induction; there is
neither magnetic dipole radiation nor the subsequent problem caused
by dipole radiation and polar motion.

\section{Verification methods}
For the non-compact stars, since their magnetic field can be measured directly by Zeeman effect of spectral lines,
several methods can be used to determine whether their magnetic fields reverse or rotate \citep{Liang09}.
Although pulsar's magnetic field cannot be measured directly,
since the MO model differs significantly from the lighthouse model, it is possible to determine which of the two models
is more reliable. Eight verification methods are presented below.

\subsection{Modulations in the double pulsar PSR J0737-3039A/B}
Each of the two pulsars in the double pulsar PSR J0737-3039A/B
system exhibits modulations near the pulse period of the other.
\citet{Freire09} and \citet{Liang14} have put forward two techniques
using these modulations to determine the
senses of rotation of the two pulsars and test the rotating lighthouse model.
The technique put forward by Liang \& Liang is entirely analogous to
the difference between the sidereal and solar day in the earth.  The direct detection of spin
will observationally validate or rule out the lighthouse model.

\subsection{Supernova (SN)1987A}
It is expected that a neutron star lies at
the center of the remnant of SN 1987A.
However, astronomers have so far not detected it. A number of possibilities
for the `missing' neutron star have been considered\citep{Arnett89}. We think another possibility
is that its period is much shorter than that estimated by lighthouse model.
If a very short period, ($<$ 0.1 ms, for example) is detected tomorrow, the lighthouse model will be ruled out.

\subsection{Effect of precession}
According to the lighthouse model, the profile and the flux are
strongly related to the precession angle in a binary system. On this
basis, \citet{JR04} predicted that the profile of PSR J0373-3039A
would evolve considerably and it would disappear entirely over a
period of 15 to 20 years. \citet{Kramer98} also predicted that PSR
B1913+16 would disappear from our sight after the year 2025.

However,the MO model think that only the polarization parameters are
strongly related to the precession, but the profile and flux are
only weakly related to the precession. After the precession of the
spin axis, the polarization characteristics will evolve
considerably, but the flux and profile will merely show a bit little
change. Therefore, the MO model makes two predictions that differ
from those of the lighthouse model:

First, the polarization characteristics could evolve considerably
even though the evolution of profile and flux were not obvious.
Second, all pulsars will not be out of our line of sight even if the
precession is very great.

After finding PSRJ0737-3039A/B, it was commonly believed that its
profile could rapidly evolve. But we once predicted that its profile
could not evolve obviously. At that time, we once discussed our
prediction with several astronomers, but no body laid stress on our
prediction. Mow, The observational results reported by
\citet{Manchester05} are clearly consistent with our prediction.

After twenty years, whether the PSR B1913+16 disappear or not will
finally determine which model is better, the lighthouse model or the
MO model.

\subsection{Effect of magnetic inclination evolution}
The calculating result which is based on the lighthouse model shows
that the evolution rate of Crab's magnetic inclination is about
$0.013^{\circ}$ per year. It is so rapid that the change of profile
must be detected after hundreds years. In contrast with the
lighthouse model, the MO model thinks that the magnetic inclination
of pulsars is neglectable, and the profile would be changeless.
After several hundred years, if the anticipative change of the
profile hadn't occurred, the lighthouse model should be ruled out.

\subsection{Relationship between the characteristics of polarization
and the direction of the spin axis} The images of PWNs taken using
the Chandra telescope (e.g. Figure \ref{Crab}) can help to estimate
the direction of the spin axis of some pulsars from the shapes of
PWNs. The relationship between the characteristics of polarization
and the direction of the spin axis is given in Figure
\ref{polarization}. This enables us to determine the relationship
between the shape of PWN and the polarization. The predictions of
the MO model are as follows. If PWN is cigar-shaped, as with PSR
J0205+6449, the radiation will be linear polarization and the
oscillation plane of the electric vector is averagely parallel with
the cigar. If the shape of PWN is round, as with PSR B0540-69, the
radiation will be low-level circular polarization, and if the shape
of PWN is elliptical, as with the Crab, the radiation will be
elliptical polarization. This relationship between polarization and
the spin axis could be used to verify the MO model.

\subsection{Sweep delay effect }
According to the lighthouse model, when the light beams sweep over
the different points on the revolving orbit of the earth, such as
the spring equinox and the autumnal equinox, a time delay should
exist. In other word, the light beams sweep over the autumnal
equinox and the spring equinox in turn, even if the distances of two
points to pulsar are absolute equal. After correcting the arrival
time to the barycenter of the solar system, there is still a very
small sine variety in the arrival time data. For example, the
maximum delay of PSR J2144-3933 is as much as 0.24 $\mu$s. It is
very difficult to measure such small time delay, but as long as the
arrival time data is accumulating continuously and calculation
accuracy is also improving, this measurement should be possible in
the future. In contrast, the delay of far pulsars can be neglected.
Therefore, correcting the revolving orbit of the earth using the
data of distant pulsars can help to detect the time delay of near
pulsars. This sweep delay is the most direct evidence of the
lighthouse model. If this sweep delay can't be observed using
methods that are highly precise enough, then the lighthouse model
can be rejected.

\subsection{Stability of the 1/60 Hz signal of Crab}
The MO model thinks that the 1/60 Hz signal of Crab \citep{Cadez01}
is caused by the spin modulation and the 1/60 Hz signal should
surpass the 1/0.033 Hz pulse signal in stability of frequency.
Therefore, making a comparison between two signals in the stability
of frequency is a Verification method.

\subsection{Interpulse in the young neutron stars}
We guess that the SGRs are young neutron stars with weaker magnetic
field. Usually, their magnetic field is oscillating, but the
radiation is too weak to be observed. Only when some substances
flied-out as the neutron stars forming return and bump the celestial
body, and gamma radiation bursts, the oscillation of weaker magnetic
field could be detected as a modulating signal. thus, we predict
that all of the radiations of SGRs have the interpulse. According to
the above discussion£¬the interpulse is closely related to the age,
but has nothing to do with look angle. Therefore, we also predict
that if a pulsar will be discovered in the SN 1987A in the future,
it must have interpulse. If these two predictions could be proved,
the guess in which the interpulse had nothing to do with the look
angle would be proved at the same time. It could only be explained
by the MO model, and the lighthouse model would be impuissant.

\section{Conclusions}
Compared to the lighthouse model, the MO model is better able to
explain all kinds of radiation characteristics of pulsars,
especially the characteristics of pulse polarization, the
microstructure of profile, the glitch, the configuration of PWN.
Although the reason why the oscillation of the magnetic field of
pulsars remains unknown, its observation characteristics lead us to
believe that the magnetic field oscillation should be the source of
pulsar radiation.

The MO model is incapable to explain the reason of magnetic field
oscillation, much less calculate the time-scale of oscillation.
Therefore, a novel theory on the origin of stellar magnetic field is
needed to support the MO model.


\begin{thebibliography}{}

\bibitem[Arnett et al.(1989)]{Arnett89}
Arnett, W. D., Supernova 1987A, ARA\&A, 27, 629,1989.





\bibitem[Babcock(1958)]{Babcock58}
Babcock, H.W., Magnetic fields of the A-type stars, ApJ.., 128, 228,
1958.

\bibitem[\protect\citeauthoryear{Freire et al.}{2009}]{Freire09}Freire P. C. C. et al.,
A new technique for timing the double pulsar system,
MNRAS, 396, 1764, 2009.


\bibitem[Hallinan et al.(2007)]{Hallinan07}
Hallinan, G., Bourke, S., Lane, C. et al., Periodic bursts of
coherent radio emission from an ultracool dwarf, ApJ.., 663L, 25,
2007.

\bibitem[Jenet \& Ransom(2004)]{JR04}
Jenet, F.A., \& Ransom, S.M., The geometry of the double-pulsar
system J0737-3039 from systematic intensity variations, Nature, 428,
919-921, 2004.

\bibitem[Kienle(1950)]{Kienle50}
Kienle, H., Zur deutung der magnetfelder der sterne, NW..., 37,
137v, 1950.

\bibitem[Kramer(1998)]{Kramer98}
Kramer, M., Determination of the geometry of the PSR B1913+16 system
by geodetic precession, ApJ.., 509, 856-860, 1998.


\bibitem[Liang \& Liang(2009)]{Liang09}
Liang, Z-X. \& Liang, Y., Six Methods for Distinguishing Rotation and
Reversal in a Stellar Magnetic Field, arXiv:0904.2522, 2009.

\bibitem[Liang et al.(2014)]{Liang14}
Liang, Z-X., Liang, Y., \& Weisberg, J. M.,  Testing the rotating lighthouse model
with the double pulsar system PSR J0737-3039A/B, MNRAS, 439, 3712, 2014.


\bibitem[Manchester(2005)]{Manchester05}
Manchester, R.N, The mean pulse profile of PSR J0737-3039A, ApJ..,
621L, 49M, 2005.

\bibitem[Moffett \& Hankins(1996)]{MH96}
Moffett, D.A., \& Hankins, T.H., Multifrequency radio observations
of the Crab pulsar, ApJ.., 468, 779-783, 1996.

\bibitem[Monaghan(1968)]{Monaghan68}
Monaghan, J.J., The precession of a rotating magnetic star, ZA...,
69, 154, 1968.




\bibitem[Trigilio et al.(2000)]{Trigilio00}
Trigilio, C., Leto, P., Leone, F., et al., Coherent radio emission from the magnetic chemically peculiar star CU Virginis, A\&A, 362, 281, 2000.





\bibitem[Popov et al.(2002)]{Popov02}
Popov, M. V., Bartel, N., Cannon, W. H. et al., Pulsar
microstructure and its quasi-periodicities with the S2 VLBI system
at a resolution of 62.5 nanoseconds, A\&A, 396, 171-187, 2002.

\bibitem[Weisskopf et al.(2000)]{Weisskopf00}
Weisskopf, M.C., Hester, J.J, Tennant, A.F. et al., Discovery of
spatial and spectral structure in the X-ray emission from the Crab
nebula, ApJ.., 536, L81, 2000.

\bibitem[\v{C}ade\v{z} et al.(2001)]{Cadez01}
\v{C}ade\v{z},A.,  Vidrih, S., Gali\v{c}i\v{c}, M. et al., Crab
pulsar photometry and the signature of free precession,
A\&A...366..930C,2001.
\end{thebibliography}
\end{document}